\documentclass[
	journal
]{IEEEtran}

\IEEEoverridecommandlockouts 
\makeatletter
\def\markboth#1#2{\def\leftmark{\@IEEEcompsoconly{\sffamily}\MakeUppercase{\protect#1}}%
\def\rightmark{\@IEEEcompsoconly{\sffamily}\MakeUppercase{\protect#2}}}
\makeatother

\usepackage[cmex10]{amsmath} 
\usepackage{amssymb}
\usepackage{bbm} 
\usepackage{bm} 
\usepackage{amsthm}
\usepackage{xspace}
\usepackage[latin1]{inputenc}
\usepackage{tikz}
\usepackage{xcolor}
\usepackage{cite}
\usepackage{graphicx}
\usepackage{subcaption}
\usepackage{balance} 
\usepackage{url}
\usepackage[normalem]{ulem}
\usepackage{lipsum}
\usepackage{booktabs} 
\usepackage{siunitx}

\theoremstyle{remark} 
\newtheorem{remark}{Remark}
\newtheorem{example}{Example}

\newcommand\xqed[1]{%
\leavevmode\unskip\penalty9999 \hbox{}\nobreak\hfill
\quad\hbox{#1}}
\newcommand\demo{\xqed{$\triangle$}}

\newcommand{\imag}{\jmath}
\newcommand{\vect}[1]{\ensuremath{\bm{#1}}}
\newcommand{\mat}[1]{\ensuremath{\bm{#1}}}

\newcommand{\Lsp}{\ensuremath{L_{\text{sp}}}}
\newcommand{\Nsp}{\ensuremath{N_\text{sp}}}
\newcommand{\Nf}{\ensuremath{N_\text{F}}}
\newcommand{\Nsym}{N_\text{sym}}
\newcommand{\OSa}{\rho_\text{a}}
\newcommand{\OSd}{\rho_\text{d}}
\newcommand{\loss}{l} 

\newcommand{\flen}{T} 

\newcommand{\nl}{\eta} 

\newcommand{\transpose}{\top}

\newcommand{\fsym}{\ensuremath{R_\text{s}}}
\newcommand{\fs}{\ensuremath{f_\text{s}}}

\newcommand{\D}{\mathrm{d}}
\newcommand{\define}{\triangleq}
\newcommand{\FFTm}{\mat{F}}
\DeclareMathOperator{\diag}{\text{diag}}

\newcommand{\NumSteps}{\ensuremath{M}}
\newcommand{\ModSteps}{\ensuremath{\ell}} 
\newcommand{\StepSize}{\ensuremath{\delta}}
\newcommand{\nlop}[1]{\ensuremath{\bm{\sigma}_{#1}}}

\newcommand{\Leff}{L_{\text{eff}}}

\newcommand{\pTX}{\mathcal{T}_{\theta}}
\newcommand{\pRX}{\mathcal{R}_{\theta}}
\newcommand{\pCH}{\mathcal{C}_{\theta}}

\definecolor{DarkGreen}{rgb}{0.0, 0.7, 0.0}
\newcommand{\NoRev}[1]{{#1}}
\newcommand{\RevA}[1]{{#1}}
\newcommand{\RevB}[1]{{#1}}
\newcommand{\RevC}[1]{{#1}}

\newcommand{\MyNum}[1]{%
\num[group-separator={,}]{#1}
}%

\begin{document}

\title{Physics-Based Deep Learning for\\ Fiber-Optic Communication
Systems}

\author{%
	\IEEEauthorblockN{%
	Christian H\"ager, \emph{Member, IEEE} and Henry D.~Pfister, \emph{Senior Member, IEEE}%
	\thanks{Parts of this paper have been presented at the Optical
	Fiber Communication Conference~(OFC), San Diego, USA, 2018,
	the International~Symposium~on Information Theory (ISIT), Vail, USA, 2018, 
	and the Information Theory Workshop (ITW), Guangzhou,
	China, 2018.}%
	\thanks{This work is part of a project that has received funding
	from the European Union's Horizon 2020 research and innovation
	programme under the Marie Sk\l{}odowska-Curie grant agreement
	No.~749798. The work of H.~D.~Pfister was supported in part by the
	National Science Foundation (NSF) under Grant No.~1609327. Any
	opinions, findings, recommendations, and conclusions expressed in
	this material are those of the authors and do not necessarily
	reflect the views of these sponsors.}%
	\thanks{C.~Häger is with the Department of Electrical Engineering,
	Chalmers University of Technology, Gothenburg, Sweden and
	H.~D.~Pfister is with the Department of Electrical and Computer
	Engineering, Duke University, Durham, USA (emails:
	christian.haeger@chalmers.se, henry.pfister@duke.edu).}%
	}%
}%

\maketitle

\vspace{-1cm}

\newif\ifShowText 
\ShowTexttrue

\newif\ifShowFull 
\ShowFullfalse

\begin{abstract}
	We propose a new machine-learning approach for fiber-optic
	communication systems whose signal propagation is governed by the
	nonlinear Schrödinger equation (NLSE). Our main observation is that
	the popular split-step method (SSM) for numerically solving the
	NLSE has essentially the same functional form as a deep multi-layer
	neural network; in both cases, one alternates linear steps and
	pointwise nonlinearities. We exploit this connection by
	parameterizing the SSM and viewing the linear steps as general
	linear functions, similar to the weight matrices in a neural
	network. The resulting physics-based machine-learning model has
	several advantages over ``black-box'' function approximators. For
	example, it allows us to examine and interpret the learned
	solutions in order to understand \emph{why} they perform well. As
	an application, low-complexity nonlinear equalization is
	considered, where the task is to efficiently invert the NLSE. This
	is commonly referred to as digital backpropagation (DBP). 
	\RevB{Rather than employing neural networks, 
	the} proposed algorithm, dubbed learned DBP (LDBP), uses \RevB{the physics-based model with} trainable
	filters in each step and its complexity is reduced by progressively
	pruning filter taps during gradient descent. Our main finding is
	that the filters can be pruned to remarkably short lengths---as few
	as $3$ taps/step---without sacrificing performance. As a result,
	the complexity can be reduced by orders of magnitude in comparison
	to prior work. By inspecting the filter responses, an
	additional theoretical justification for the learned parameter
	configurations is provided. Our work illustrates that combining
	data-driven optimization with existing domain knowledge can
	generate new insights into old communications problems.
\end{abstract}

\begin{IEEEkeywords}
	Deep neural networks, 
	digital backpropagation, 
	machine learning,
	nonlinear equalization, 
	nonlinear interference mitigation, 
	physics-based deep learning, 
	split-step method
\end{IEEEkeywords}

\section{Introduction}
\label{sec:introduction}

Rapid improvements in machine learning over the past decade are
beginning to have far-reaching effects. In particular the use of deep
learning to progressively process raw input data into a hierarchy of
intermediate signal (or feature) vectors has led to breakthroughs in
many research fields such as computer vision or natural language
processing \cite{LeCun2015, Goodfellow2016}. The success of deep
learning has also fuelled a resurgence of interest in machine-learning
techniques for communication systems assuming a wide variety of
channels and applications \cite{Zibar2016, Wang2017, Simeone2018,
Khan2019}. In this paper, we are interested in the question to what
extend supervised learning, and in particular deep learning, can
improve physical-layer communication over optical fiber.

The traditional application of supervised learning to physical-layer
communication replaces an individual digital signal processing (DSP)
block (e.g., equalization or decoding) by a neural network (NN) with
the aim of learning better-performing (or less complex) algorithms
through data-driven optimization \cite{Ibnkahla2000}. More generally,
one can treat the entire design of a communication system as an
end-to-end reconstruction task and jointly optimize transmitter and
receiver NNs \cite{OShea2017}. Both traditional \cite{Shen2011,
Jarajreh2015, Gaiarin2016, Ahmad2016, Estaran2016, Eriksson2017,
Kamalov2018} and end-to-end learning \cite{Shen2018ecoc, Jones2018,
Karanov2018, Karanov2019, Jones2019, Song2020} have received
considerable attention for optical-fiber systems. However, the
reliance on NNs as universal (but sometimes poorly understood)
function approximators makes it difficult to incorporate existing
domain knowledge or interpret the obtained solutions.

Rather than relying on generic NNs, a different approach is to start
from an existing model or model-based algorithm and extensively
parameterize it. For iterative algorithms, this can be done by
``unfolding'' the iterations which gives an equivalent feed-forward
computation graph with multiple layers. This methodology has been
applied, for example, in the context of decoding linear codes via
belief propagation\cite{Nachmani2018}, MIMO detection
\cite{Samuel2017}, and sparse signal recovery \cite{Gregor2010,
Borgerding2016}. An overview of applications for communication systems
can be found in \cite{Balatsoukas-Stimming2019}. Besides unfolding
iterative algorithms, there also exist various domain-specific
approaches, see, e.g., \cite{OShea2017, He2019}. As an example, the
authors in \cite{OShea2017} propose so-called radio-transformer
networks (RTNs) that apply predefined correction algorithms to the
signal. A separate NN can then be used to provide parameter estimates
to the individual RTNs.

The main contribution in this paper is a novel machine-learning
approach for fiber-optic systems where signal propagation is governed
by the nonlinear Schrödinger equation (NLSE)
\cite[p.~40]{Agrawal2006}. Our approach is based on the popular
split-step method (SSM) for numerically solving the NLSE. The main
idea is to exploit the fact that the SSM has essentially the same
functional form as a multi-layer NN; in both cases, one alternates
linear steps and pointwise nonlinearities. The linear SSM steps
correspond to linear transmission effects such as chromatic dispersion
(CD). By parameterizing these steps and viewing them as general linear
functions, one obtains a parameterized multi-layer model. Compared to
standard ``black-box'' NNs, we show that this physics-based approach
has several compelling advantages: it leads to clear hyperparameter
choices (such as the number of layers/steps); it provides good
initializations for a gradient-based optimization; and it allows us to
easily inspect and interpret the learned solutions in order to
understand \emph{why} they work well, thereby providing significant
insight into the problem.

The approach in this paper and some of the results have previously
appeared in various conference papers \cite{Haeger2018ofc,
Haeger2018isit, Lian2018itw}. The main purpose of this paper is to
provide a comprehensive treatment of the approach, subsuming
\cite{Haeger2018ofc, Haeger2018isit, Lian2018itw}. This paper also
contains several additional contributions: 
\RevC{\begin{itemize}
	\item We significantly enlarge our simulation setup compared to \cite{Haeger2018ofc, Haeger2018isit, Lian2018itw} and provide a thorough numerical investigation of the proposed approach. 
This includes a characterization of the generalization capabilities of the trained models when varying the transmit power and the employed modulation format, as well as a comparison between different parameter-initialization schemes. 

	\item We conduct an in-depth investigation of the performance--complexity trade-off of the proposed scheme. 
	In particular, we demonstrate that the model complexity in terms of the per-step filter response lengths can be close to the theoretically expected minimum lengths according to the memory introduced by CD (Fig.~\ref{fig:required_length}). 
	We also extend this investigation to higher baud-rate ($32$ Gbaud) signals, which allows us to directly compare to prior work in \cite{Martins2018}, showing significant complexity advantages.

	\item We also study wavelength division multiplexing (WDM) transmission, where we demonstrate that the resulting relaxed accuracy requirements for solving the NLSE provide an opportunity to further reduce the complexity of the proposed scheme. 

	\item Lastly, we show that our approach generalizes to the so-called enhanced (or filtered) SSM \cite{Du2010, Secondini2016} where an additional filtering step is included in the nonlinear steps.

\end{itemize}%
}%

We also note that since the publication
of the original paper \cite{Haeger2018ofc}, multiple extensions have
appeared in the literature to address issues such as hardware
implementation \cite{Fougstedt2018ecoc}, subband processing
\cite{Haeger2018ecoc}, polarization multiplexing \cite{Haeger2020ofc,
Bitachon2020}, training in the presence of practical impairments
(e.g., phase noise) and experimental demonstration \cite{Bitachon2020},\NoRev{\cite{Bitachon2020b}},\cite{Sillekens2019, Oliari2020},\NoRev{\cite{Fan2020}}.

The remainder of the paper is structured as follows. In
Sec.~\ref{sec:supervised_ml}, we give a brief introduction to
supervised machine learning and multi-layer NNs. The proposed approach
is then discussed in detail in Sec.~\ref{sec:model-based}.
Sec.~\ref{sec:ldbp} describes its application to nonlinear
equalization via digital backpropagation (DBP). Numerical results are
presented in Sec.~\ref{sec:results}, some of which are then further
examined and interpreted in Sec.~\ref{sec:examining}. Finally,
Sec.~\ref{sec:conclusion} concludes the paper. 
\RevA{A list of symbols used in this paper can be found in Table~\ref{tab:parameters} below.}

\section{Supervised Learning with Neural Networks}
\label{sec:supervised_ml}

We start in this section by briefly reviewing the standard supervised
learning setting for feed-forward NNs. It is important to stress,
however, that we do not directly use conventional NNs in this work.
Instead, their underlying mathematical structure serves as the main
motivation for the proposed approach described in the next section.

\subsection{Neural Networks}

\begin{figure}[t]
	\centering
		\includegraphics{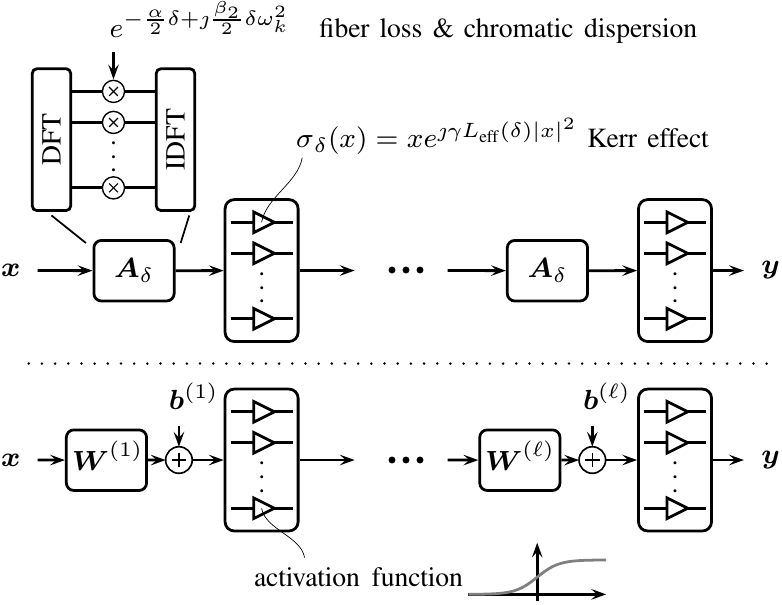}
	\caption{Block diagram of the split-step method to numerically
	solve the nonlinear Schrödinger equation (top) and the canonical
	model of a deep feed-forward neural network (bottom).}
	\label{fig:ssfm}
\end{figure}

A deep feed-forward NN with $\ell$ layers defines a parameterized
mapping $\RevB{\vect{y}} = \vect{f}_\theta(\RevB{\vect{x}})$ where the input vector
$\RevB{\vect{x}} \in \RevB{\mathcal{X}}$ is mapped to the output vector $\RevB{\vect{y}}
\in \RevB{\mathcal{Y}}$ through a series of affine transformations and
pointwise nonlinearities \cite{LeCun2015}, \cite[Eq.~(6)]{Lin2017}.
Here, $\theta$ is a parameter vector and $\RevB{\mathcal{X}}$ and
$\RevB{\mathcal{Y}}$ denote the input and output alphabet, respectively. The
affine transformations are defined by 
\begin{align}
	\RevB{\vect{\tilde{x}}^{(i)}} = \mat{W}^{(i)} \RevB{\vect{x}^{(i-1)}} +
	\vect{b}^{(i)}, \qquad i = 1, 2, \dots, \ell,
\end{align}
where $\RevB{\vect{x}} \define \RevB{\vect{x}^{(0)}}$ is the input to the first
layer, $\mat{W}^{(i)}$ is a weight matrix, and $\vect{b}^{(i)}$ is a
bias vector. The nonlinearities are defined by $\RevB{\vect{x}^{(i)}} =
\NoRev{\bm{\sigma}}(\RevB{\vect{\tilde{x}}^{(i)}})$, where $\NoRev{\bm{\sigma}}$ refers to the
element-wise application of some nonlinear activation function $\NoRev{\sigma}$.
Common choices for the activation function include
$\NoRev{\sigma}(z)=\max\{0,z\}$, $\NoRev{\sigma}(z)=\tanh(z)$, and $\NoRev{\sigma}(z)=1/(1+e^{-z})$.
The final output of the NN after the $\ell$-th layer is $\RevB{\vect{y}}
\define \RevB{\vect{x}^{(\ell)}}$. The block diagram illustrating the entire
NN mapping is shown in the bottom part of Fig.~\ref{fig:ssfm}.

The parameter vector $\theta$ encapsulates all elements in the weight
matrices and bias vectors, where we write $\theta = \{\vect{W}^{(1)},
\dots, \vect{W}^{(m)}, \vect{b}^{(1)},\dots,\vect{b}^{(m)}\}$ with
some abuse of notation. NNs are universal function approximators, in
the sense that they can approximate any desired function arbitrarily
well by properly choosing the parameters $\theta$ and assuming a
sufficiently large network architecture \cite{Hornik1989}.

\subsection{Supervised Learning}
\label{sec:supervised_l}

In a supervised learning setting, one has a training set $S\subset
\RevB{\mathcal{X}} \times \RevB{\mathcal{Y}}$ containing a list of input--output
pairs that implicitly define a desired mapping. Then, training
proceeds by minimizing the empirical training loss $\mathcal{L}_S
(\theta)$, where the empirical loss $\mathcal{L}_A (\theta)$ for a
finite set $A\subset \RevB{\mathcal{X}}\times \RevB{\mathcal{Y}}$ of input--output
pairs is defined by
\begin{align}
	\label{eq:loss}
	\mathcal{L}_A (\theta) \triangleq \frac{1}{|A|}
	\sum_{(\RevB{\vect{x}},\RevB{\vect{y}})\in A} \loss \big( f_\theta(\RevB{\vect{x}}),
	\RevB{\vect{y}} \big) 
\end{align}
and $\loss(\RevB{\hat{\vect{y}}}, \RevB{\vect{y}})$ is the per-example loss
function
associated with returning the output $\RevB{\hat{\vect{y}}} =
\vect{f}_\theta(\RevB{\vect{x}})$ when $\RevB{\vect{y}}$ is correct. When
the training set is large, the parameter vector $\theta$ is commonly
optimized by using a variant of stochastic gradient descent (SGD). In
particular, standard mini-batch SGD uses the parameter update
\begin{align}
	\theta_{t+1} = \theta_t - \NoRev{\mu} \nabla \mathcal{L}_{B_t} (\theta_t),  
\end{align}
where $\NoRev{\mu}$ is the learning rate and $B_t \subseteq S$ is the
mini-batch used in the $t$-th step. Typically, $B_t$ is chosen to be a
random subset of $S$ with some fixed size that matches available
computational resources, e.g., graphical processing units (GPUs).

Supervised learning is not restricted to NNs and learning algorithms
such as SGD can be applied to other function classes as well. Indeed,
in this paper, we do not use NNs, but instead exploit the underlying
signal-propagation dynamics in optical fiber to construct a suitable
model $\vect{f}_\theta$.

\section{Physics-Based Machine-Learning Models} 
\label{sec:model-based}

\begin{figure}[t]
	\centering
	\includegraphics{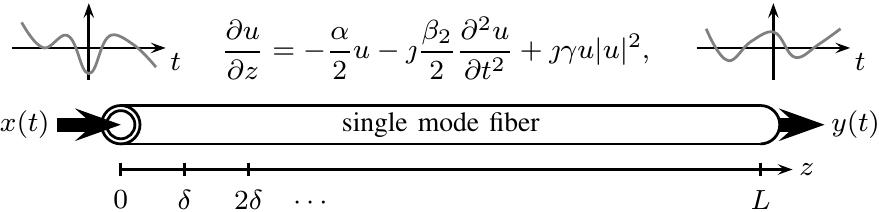}
	\caption{Conceptual signal evolution in a single-mode fiber. The
	nonlinear Schrödinger equation implicitly describes the
	relationship between the input signal $x(t) = u(z=0,t)$ and the
	output signal $y(t) = u(z=L,t)$.}
	\vspace*{-0.2cm}
	\label{fig:fiber}
\end{figure}

Consider a complex-valued baseband signal $x(t)$, which is transmitted
over an optical fiber as illustrated in Fig.~\ref{fig:fiber}. The
signal propagates according to the NLSE \cite[p.~40]{Agrawal2006}
\begin{align}
	\label{eq:nlse}
	\frac{\partial u(z,t)}{\partial z} = \RevB{-\frac{\alpha}{2} u(z,t) }  - \imag
	\frac{\beta_2}{2} \frac{\partial^2 }{\partial t^2} u(z,t)+ \imag
	\gamma |u(z,t)|^2 u(z,t),
\end{align}
where $u(z=0, t) \define x(t)$, \RevB{$\alpha$ is the loss parameter,} $\beta_2$ is the CD coefficient, and $\gamma$ is the nonlinear Kerr parameter. 
The signal after propagation distance $L$ is denoted by $y(t) \define u(z=L, t)$. \RevB{Note that \eqref{eq:nlse} covers single-polarization systems. The extension to dual-polarization transmission is discussed in Sec.~\ref{sec:dual_polarization}.}

\subsection{The Split-Step Method}
\label{sec:ssm}

In general, \eqref{eq:nlse} does not admit a closed-form solution and
must be solved using numerical methods. One of the most popular
methods is the SSM which is typically implemented via block-wise
processing of sampled (i.e., discrete-time) waveforms. To that end,
assume that the signals $x(t)$ and $y(t)$ are sampled at $t = k / \fs
$ to give sequences of samples $\{x_k\}_{k \in \mathbb{Z}}$ and
$\{y_k\}_{k \in \mathbb{Z}}$, respectively. We further collect $n$
consecutive samples into the respective vectors $\vect{x} = (x_1,
\dots, x_n)^\transpose$ and $\vect{y} = (y_1, \dots, y_n)^\transpose$,
where $\vect{x}, \vect{y} \in \mathbb{C}^n$. In order to derive the
SSM, it is now instructive to consider the time-discretized NLSE
\begin{align}
	\label{eq:discretized_nlse}
	\frac{\D \vect{u}(z)}{\D z} 
	= \RevB{-\frac{\alpha}{2} \vect{u}(z)} +  \mat{A} \vect{u}(z) +
	\imag \gamma \bm{\rho}(\vect{u}(z)), 
\end{align}
where $\mat{A} = \FFTm^{-1} \diag(H_1, \dots, H_n) \FFTm$, $\FFTm$ is
the $n\times n$ discrete Fourier transform (DFT) matrix, $H_k = \imag
\frac{\beta_2}{2} \omega_k^2$, $\omega_k = 2 \pi f_k$ is the $k$-th
DFT angular frequency (i.e., $f_k/\fs = (k-1)/n$ if $k < n/2$ and
$f_k/\fs = (k-1-n)/n$ if $k \geq n/2$), and $\bm{\rho} : \mathbb{C}^n
\to \mathbb{C}^n$ is defined as the element-wise application of
$\rho(x) = x |x|^2$. Eq.~\eqref{eq:discretized_nlse} is a first-order
ordinary differential equation with a vector-valued function
$\vect{u}(z)$, where $\vect{u}_0 \define \vect{u}(0) = \vect{x}$ is
the initial boundary condition. Next, the fiber is conceptually
divided into $\NumSteps$ segments of lengths $\delta_1, \ldots,
\delta_M$ such that $\sum_{i=1}^{M} \delta_i = L$. It is then assumed
that for sufficiently small $\delta_i$, the effects stemming from the
\RevB{linear and nonlinear} terms on the right hand side of \eqref{eq:discretized_nlse} can be
separated. More precisely, for $\gamma = 0$,
\eqref{eq:discretized_nlse} is linear with solution $\vect{u}(z) =
\mat{A}_z \vect{u}_0$, where $\mat{A}_z \define \RevB{e^{-\frac{\alpha}{2}z}} e^{z \mat{A}}$ \RevB{and} $e^{z \mat{A}} =
\FFTm^{-1} \diag(e^{z H_1}, \dots, e^{z H_n}) \FFTm$. For \RevB{$\alpha =0$ and} $\beta_2 =
0$, one may verify that the solution is $\vect{u}(z) =
\nlop{z}(\vect{u}_0)$, where $\nlop{z} : \mathbb{C}^n \to
\mathbb{C}^n$ is the element-wise application of $\sigma_z(x) = x
e^{\imag \gamma z |x|^2 }$. Alternating between these two operators
for $z=\delta_i$ leads to the SSM
\begin{align}
	\label{eq:symmetric_ssm}
	\vect{u}_{i} = 
	\nlop{\StepSize_i}\left(
	\mat{A}_{\StepSize_i} \vect{u}_{i-1}\right),
	\qquad i = 1, 2, \dots, \NumSteps,
\end{align}
where $\vect{u}_i$ serves as an estimate $\vect{u}_{i} \approx
\vect{u}(\delta_1 + \dots + \delta_i)$. 
\RevB{We note that it is typically more accurate to also consider the effect of the loss term in the nonlinear step by using $\sigma_z(x) = x
e^{\imag \gamma \RevB{\Leff(z)} |x|^2 }$ where $\Leff(z) = (1 - e^{-\alpha z})/\alpha $ is the so-called effective nonlinear length}. 
The corresponding block
diagram for $\delta_1 = \cdots = \delta_M = \delta$ is shown in the
top of Fig.~\ref{fig:ssfm}.

The degree to which ${\vect{u}_M}$ constitutes a good approximation of
the sampled signal vector $\vect{y} = \vect{u}(L)$ (and therefore the
true waveform $y(t)$ at distance $L$) is now a question of choosing
$M$, $\fs$, and $n$. In practice, the sampling frequency $\fs$ and the
number of steps $M$ are chosen to ensure sufficient temporal and
spatial resolution, respectively. The block length $n$ can be chosen
such that the overhead is minimized in overlap-and-save techniques for
continuous data transmission.

\subsection{Parameterizing the Split-Step Method}

From the side-by-side comparison in the top and bottom of
Fig.~\ref{fig:ssfm}, it is evident that the SSM has essentially the
same functional form as a deep feed-forward NN; in both cases, one
alternates between linear (or affine) steps and simple pointwise
nonlinearities. The main idea in this paper is to exploit this
observation by (i) appropriately parameterizing the SSM, and (ii)
applying the obtained parameterized model instead of a generic NN. 

\begin{remark}
	Conventional deep-learning tasks such as speech or object
	recognition are seemingly unrelated to nonlinear signal propagation
	over optical fiber. One may therefore wonder if the similarity to
	the SSM is merely a coincidence. In that regard, some authors argue
	that deep NNs perform well because their functional form matches
	the hierarchical or Markovian structure that is present in most
	real-world data \cite{Lin2017}. Indeed, the SSM can be seen as a
	practical example where such a structure arises, i.e., by
	decomposing the physical process described by \eqref{eq:nlse} into
	a hierarchy of elementary steps. Similar observations were made in
	\cite{Raissi2019}, where the training of deep NNs is augmented by
	penalizing non-physical solutions.
\end{remark}

Our parameterization of the SSM is directly inspired by the functional
form of NNs. In particular, instead of using the matrix
$\mat{A}_{\delta_i}$ in each of the $M$ steps, we propose to fully
parameterize all linear steps by assuming general matrices
$\mat{A}^{(1)}, \dots, \mat{A}^{(\ModSteps)} \in \mathbb{C}^{n \times
n}$, similar to the weight matrices in a NN, where $\ell = M$ refers
to the number of model steps (or layers). By combining the
parameterized linear steps and the standard nonlinearities in the SSM,
one obtains a parameterized version of the SSM according to
\begin{align}
	\label{eq:model}
	\vect{f}_\theta(\vect{x}) = \nlop{\ModSteps}(
	\mat{A}^{(\ModSteps)} \dots \nlop{1} (\mat{A}^{(1)} \vect{x})),
\end{align}
where $\theta = \{\mat{A}^{(1)}, \dots, \mat{A}^{(\ModSteps)}\}$
collects all tunable parameters and we have defined $\nlop{i} \define
\nlop{\delta_i}$.

\begin{remark}
	\label{rm:nonlinear_parameterization}
	The nonlinear operators can also be parameterized
	\cite{Haeger2018ofc}, e.g., by introducing scaling factors
	$\nl^{(1)}, \dots, \nl^{(\ModSteps)} \in \mathbb{R}$ according to
	$\sigma_i(x) = x e^{\imag \nl^{(i)} \gamma \RevB{\Leff(\delta_i)} |x|^2}$. 
	\RevA{However, it can be shown that this may lead to an overparameterized model, see Remark 4 below.} In this paper, we do not consider this parameterization and assume
	fixed nonlinearities. 
\end{remark}

\begin{figure}[t]
	\centering
		\includegraphics{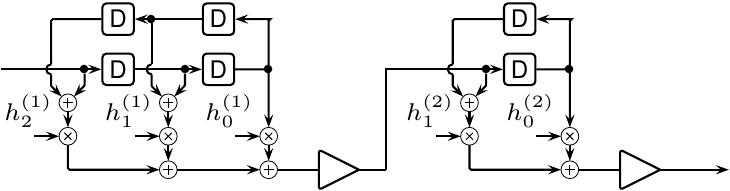}
		\caption{DSP structure with folded FIR filters for $\ModSteps = 2$,
		$\flen_1 = 5$, and $\flen_2 = 3$. Triangles represent the
		(pointwise) nonlinear steps.}
	\label{fig:filter}
\end{figure}

The above parameterization has the drawback that the number of
parameters grows quadratically with the block length $n$. A more
practical parameterization is obtained by constraining each matrix
$\mat{A}^{(i)}$ to an equivalent (circular) convolution with a filter.
In this paper, we restrict ourselves to finite impulse response (FIR)
filters which are further constrained to be symmetric. In this case,
all matrix rows are circularly-shifted versions of $(h^{(i)}_{K_i},
\dots, h^{(i)}_{1}, h^{(i)}_{0}, h^{(i)}_{1}, \dots, h^{(i)}_{K_i}, 0,
\dots, 0)$, where $h^{(i)}_j \in \mathbb{C}$, $j = -K_i, \dots, K_i$,
are the filter coefficients and $\flen_i = 2 K_i + 1$ is the filter
length. After collecting all tunable coefficients in a vector
$\vect{h}^{(i)} \define (h^{(i)}_{0}, h^{(i)}_{1}, \dots,
h^{(i)}_{K_i})$, the new parameter vector for the entire model 
is $\theta = \{\vect{h}^{(1)}, \dots, \vect{h}^{(\ModSteps)}\}$.
This reduces the number of real parameters per weight matrix from $2
n^2$ to $2 (K_i + 1)$, where typically $K_i \ll n$. This restriction
also implies that the resulting model is fully compatible with a
time-domain filter implementation \cite{Fougstedt2017}, where
symmetric coefficients can be exploited by using a folded
implementation. An example of the resulting DSP structure for
$\ModSteps = 2$ steps with $\flen_1 = 5$ and $\flen_2 = 3$ is shown in
Fig.~\ref{fig:filter}. 

\begin{remark}
	\label{rmk:special_cases}
It is worth pointing out that several problems that have been
previously studied for the standard SSM can be regarded as special
cases of the proposed model. For example, it is well known that
non-uniform step sizes can improve accuracy \cite{Bosco2000}, which
leads to the problem of finding an optimized step-size distribution.
Another common problem is to properly adjust the nonlinearity
parameter in each step (i.e., using $\gamma'$ instead of the actual
fiber parameter $\gamma$) which is sometimes referred to as
``placement'' optimization of the nonlinear operator \cite{Du2010}.
Since our model implements general linear steps, a step-size
optimization can be thought of as being implicitly performed.
Moreover, it can be shown that adjusting the nonlinearity parameters
is equivalent to rescaling the filter coefficients in the linear
steps, thereby adjusting the filter gain. Thus, a joint nonlinear
operator placement can also be regarded as a special case. Indeed, the
proposed model is more general and can learn and apply arbitrary
filter shapes to the signal in a distributed fashion. This includes,
e.g., low-pass filters, which have recently been shown to improve
simulation accuracy \cite{Li2020}. 
\end{remark}

\subsection{Generalization to other Split-Step Methods}
\label{sec:generalization}

We now discuss several generalizations of the model \eqref{eq:model},
focusing on other variants of the SSM. 

The numerical method described in Sec.~\ref{sec:ssm} corresponds to
the so-called \emph{asymmetric} version of the SSM. In practice, a
\emph{symmetric} version is typically used because it reduces the
approximation error from $\mathcal{O}(M^{-1})$ to
$\mathcal{O}(M^{-2})$ at almost no extra cost \cite{Agrawal2006}. The
symmetric SSM is defined by
\begin{align}
	\label{eq:symmetric_ssm}
	\vect{u}_{i} = \mat{A}_{\StepSize_i /2} \nlop{\StepSize_i}\left(
	\mat{A}_{\StepSize_i /2} \vect{u}_{i-1}\right), \qquad i = 1, 2,
	\dots, \NumSteps.
\end{align}
The model $\vect{f}_\theta$ in \eqref{eq:model} can also be based on
parameterizing the symmetric SSM, which was done in
\cite{Haeger2018ofc, Fougstedt2018ecoc}. In that case, the general
form in \eqref{eq:model} remains valid. The main difference is that
$\ModSteps = M+1$, i.e., an $\ModSteps$-step model is based on the
symmetric SSM with $\ModSteps-1$ steps. The last model step only
corresponds to a linear ``half-step'' without application of the
nonlinear Kerr operator. This can be achieved by setting
$\delta_\ModSteps = 0$, which gives $\sigma_{\ModSteps}(x) = x
e^{\imag \gamma \RevB{\Leff(\delta_\ModSteps)} |x|^2} = x$ and thus
$\nlop{\ModSteps}(\vect{x}) = \vect{x}$. 

\begin{remark}
	\label{rm:overparameterization}
	For models based on the symmetric SSM, one can also show that
	introducing scaling factors in the nonlinear steps (as suggested in
	Remark \ref{rm:nonlinear_parameterization}) leads to an
	overparameterization, in the sense that the parameterization of the
	nonlinear operators via $\nl^{(1)}, \dots, \nl^{(\ModSteps)}$ is
	redundant. More precisely, consider the model $\vect{f}_\theta$ in
	\eqref{eq:model} with parameterized nonlinear steps, i.e., $\theta = \{\vect{h}^{(1)}, \dots,
	\vect{h}^{(\ModSteps)}, \nl^{(1)}, \dots, \nl^{(\ModSteps)} \}$ and
	assume that $\nl^{(\ModSteps)} = 0$. Then, there always exists
	$\tilde{\theta} = \{\tilde{\vect{h}}^{(1)}, \dots,
	\tilde{\vect{h}}^{(\ModSteps)}, \tilde{\nl}^{(1)}, \dots,
	\tilde{\nl}^{(\ModSteps)} \}$ with $\tilde{\nl}^{(1)} = \cdots =
	\tilde{\nl}^{(\ModSteps-1)} = 1$ and $\tilde{\nl}^{(\ModSteps)} =
	0$, such that $\vect{f}_\theta(\vect{x}) =
	\vect{f}_{\tilde{\theta}}(\vect{x})$ for all $\vect{x} \in
	\mathbb{C}^n$. 
	The proof is based on renormalizing the intermediate signals prior
	to the nonlinearities and absorbing the normalization constants
	into the filter coefficients. 
\end{remark}

Besides using the symmetric SSM, it can also be beneficial to filter
the squared modulus of the signal prior to applying the nonlinear
phase shift \cite{Du2010}. In the context of nonlinearity
compensation, this approach is known as filtered (or low-pass
filtered) DBP and variations of this idea have been explored in, e.g.,
\cite{Li2011, Rafique2011a, Secondini2016}. We adopt the terminology
from \cite{Secondini2016} and refer to the corresponding numerical
method as the enhanced SSM (ESSM). The main difference with respect to
the standard SSM is that the nonlinear steps are modified according to
\begin{align}
	\label{eq:nl_filtered}
	[\nlop{z}(\vect{x})]_j = x_j e^{\imag \gamma \RevB{\Leff(z)}
	\sum_{k=-\kappa}^\kappa
	\nl_k
	|x_{j-k}|^2 },
\end{align}
where $[\vect{a}]_j = a_j$ returns the $j$-th entry of $\vect{a}$ and
the indexing $x_{j-k}$ is interpreted modulo $n$. Compared to standard
nonlinearities, the phase shift at time $j$ now also depends on the
signal at other time instances through the filter $\vect{\nl} =
(\nl_{-\kappa}, \dots, \nl_0, \dots, \nl_\kappa)$, where the filter
taps are assumed to be real-valued. We propose to generalize the ESSM
by again parameterizing \emph{all} steps according to
\begin{align}
	\label{eq:nl_filtered}
	[\nlop{i}(\vect{x})]_j = x_j e^{\imag \gamma \RevB{\Leff(z)}
	\sum_{k=-\kappa_i}^{\kappa_i}
	\nl_k^{(i)}
	|x_{j-k}|^2 }, 
\end{align}
where $\vect{\nl}^{(i)} = (\nl_{-{\kappa_i}}^{(i)}, \dots,
\nl_0^{(i)}, \dots, \nl_{\kappa_i}^{(i)})$, $i = 1, \dots, \ModSteps$, are
different filters in each step. The optimization parameters then
become $\theta = \{\vect{h}^{(1)}, \dots, \vect{h}^{(\NumSteps)},
\vect{\nl}^{(1)}, \dots, \vect{\nl}^{(\NumSteps)}\}$. We demonstrate
in Sec.~\ref{sec:filtered_dbp} that such a parameterization indeed
leads to improvements compared to prior work that uses the same filter
$\vect{\nl}$ in each nonlinear step. 

Lastly, there are other versions of the SSM where our approach
could be applied. One example is the iterative SSM described in
\cite[Sec.~2.4.1]{Agrawal2006}. This method falls into the class of
predictor--corrector methods, where the corrector step is iterative.
While we have not explored the parameterization of the iterative SSM
in this work, the general idea is to ``unroll'' the iterations in
order to obtain a proper feed-forward computation graph that can
again be parameterized.

\begin{remark}
	Finite-difference methods have also be used extensively to solve
	the NLSE. In fact, they can be more computationally efficient than
	the SSM in some applications \cite[2.4.2]{Agrawal2006}. However, to
	the best of our knowledge, finite-difference methods have not been
	studied for real-time applications such as DBP. One reason for this
	might be that many methods that show good performance are implicit,
	i.e., they require solving a system of equations at each step. This
	makes it challenging to satisfy a real-time constraint. 
\end{remark}

\subsection{\RevB{Generalization to Dual-Polarization Systems}}
\label{sec:dual_polarization}

\RevB{Our approach also generalizes beyond the standard NLSE in \eqref{eq:nlse} to the case of \emph{coupled} NLSEs. These arise, for example, in the context of subband processing, where we refer the interested reader to \cite{Haeger2018ecoc} for more details. 
Another important example is the generalization to dual-polarization systems. In this case, the signal evolution is described by a set of coupled NLSEs that takes into account the interactions between the two (degenerate) polarization modes. 
In birefringent fibers where polarization states change rapidly along the link, an appropriate approximation is given by the so-called Manakov-PMD equation \cite{Wai1991}. 
Parameterizing the SSM for the Manakov-PMD equation was previously done in \cite{Haeger2020ofc} (with early work briefly described in \cite{Haeger2019ecoc}) and also recently in \cite{Fan2020}. 
In general, however, these generalizations lead to nontrivial choices for the specific parameterizations of the linear and nonlinear steps, and a detailed discussion is beyond the scope of this paper.
Note for example that the parameterizations described in \cite{Haeger2020ofc} and \cite{Fan2020} are different and more research is needed to properly compare the performance of the resulting systems.

}%

\section{Application to Nonlinear Interference Mitigation: Learned
Digital Backpropagation}
\label{sec:ldbp}

As an application, we consider nonlinear equalization via
receiver-side digital backpropagation (DBP) \cite{Li2008, Mateo2008,
Ip2008, Millar2010}, with an emphasis on a low-complexity hardware-efficient
implementation. DBP exploits the fact that, in the absence of noise,
the transmitted signal can be recovered by solving an initial value
problem (IVP) using the received signal as a boundary condition
\cite{Pare1996}. In practice, the received signal first passes through
an analog-to-digital converter and the IVP can then be solved
approximately via receiver DSP. Here, we apply the parameterized SSM
introduced in the last section and show how to optimize it using
machine-learning tools. We refer to the resulting approach as learned
DBP (LDBP).

A major issue with DBP is the large computational burden associated
with a real-time DSP implementation. Thus, various techniques have
been proposed to reduce its complexity \cite{Ip2008, Tanimura2009,
Du2010, Xie2010, Tao2011, Li2011, Shen2011, Rafique2011a, Liang2014,
Jarajreh2015, Giacoumidis2015, Secondini2016, Fougstedt2017,
Fougstedt2017b, Nakashima2017, Fougstedt2018ptl}. In essence, the task
is to approximate the solution of a partial differential equation
using as few computational resources as possible.  We approach this
problem from a machine-learning perspective by applying model
compression, which is commonly used to reduce the size of NNs
\cite{Lecun1989, Han2016}. As explained in detail below, we use a
pruning-based approach, where the filters in the LDBP model are
progressively shortened during SGD.

\begin{figure*}[t]
	\centering
		\includegraphics{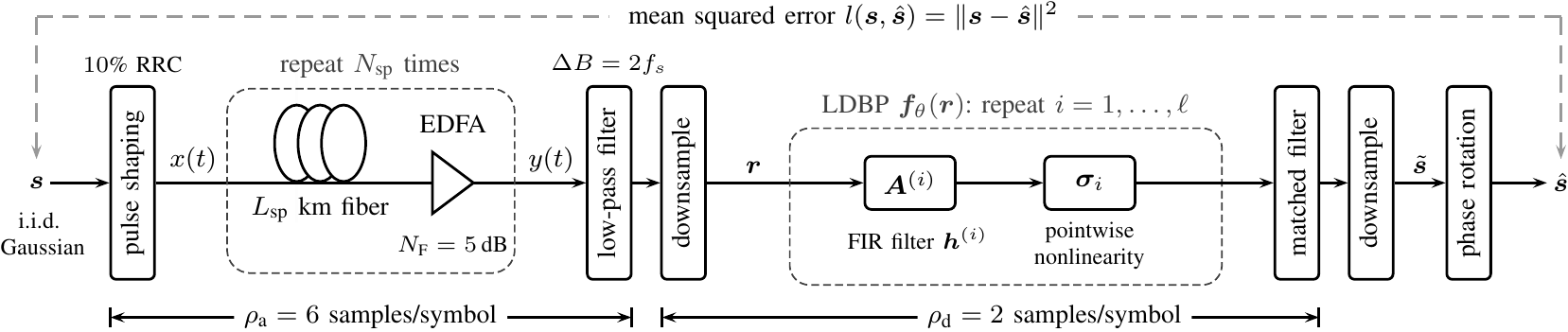}
	\caption{Block diagram showing the entire end-to-end system model. }
	\label{fig:block_diagram}
\end{figure*}

\subsection{System Model}

A block diagram of the system model is shown in
Fig.~\ref{fig:block_diagram}. The transmitted signal is given by 
\begin{align}
	x(t) = \sqrt{P} \sum_{k=-\infty}^\infty s_{k}
	p\left(t-\frac{k}{\RevA{\fsym}}\right),
\end{align}
where $P$ is the signal power, $\{s_k\}_{k \in \mathbb{Z}}$ is a
sequence of complex symbols, $p(t)$ is \RevB{the impulse response of a root-raised cosine (RRC) filter with $10\%$ roll-off factor}, and $\RevA{\fsym}$ is the baud rate. The symbol sequence is assumed to
be periodic consisting of repeated versions of the vector $\vect{s} =
(s_1, \dots, s_{\Nsym}) \in \mathbb{C}^{\Nsym}$. The symbols in
$\vect{s}$ are independent and identically distributed (i.i.d.) \RevB{circularly-symmetric complex Gaussian}, and
further normalized such that $\mathbb{E}[|S_i|^2] = 1$. The signal
$x(t)$ is assumed to be transmitted over an optical link consisting of
$\Nsp$ spans of standard single-mode fiber. Each span has length
$\Lsp$ and an erbium-doped fiber amplifier (EDFA) with noise figure
$\Nf$ is inserted after each span to compensate exactly for the
incurred signal attenuation. Each EDFA introduces white Gaussian noise
with power spectral density $(e^{\alpha \Lsp}-1) h \nu_s n_\text{sp} $
\cite[Eq.~(54)]{Essiambre2010}, where $h$ is Planck's constant,
$\nu_s$ is the optical carrier frequency, and $n_\text{sp} = \Nf (2 (1
- e^{-\alpha \Lsp}))^{-1}$ is the spontaneous emission factor.  After
propagation distance $L = \Nsp \Lsp$, the received signal $y(t)$ is
low-pass filtered using an ideal brick-wall filter with bandwidth
$\Delta B = 2 \fs$ and sampled at $t = k / \fs$ to give a sequence of
received samples $\{r_k\}_{k \in \mathbb{Z}}$, where $\OSd = \fs /
\RevA{\fsym}$ is the digital oversampling factor.  The first $n = \OSd \Nsym$
samples are collected into the observation vector $\vect{r} = (r_1,
\dots, r_{n}) \in \mathbb{C}^{n}$. Forward propagation is simulated
using $\OSa > \OSd$ samples/symbol.

The goal is to recover $\vect{s}$ from the observation $\vect{r}$. To
that end, we use a receiver DSP chain that consists of three blocks:
(i) LDBP, (ii) a digital matched filter \RevB{(i.e., RRC filter with $10\%$ roll-off factor)} followed by downsampling, and
(iii) a phase-offset correction. LDBP alternates linear and nonlinear
steps according to \eqref{eq:model}, as shown in
Fig.~\ref{fig:block_diagram}. 
The phase-offset correction is \RevB{performed} according to
$\hat{\vect{s}} = \tilde{\vect{s}} e^{-\imag \hat{\phi}}$, where
$\hat{\phi} = \arg ( \vect{s}^\dag \tilde{\vect{s}}) $ and
$\tilde{\vect{s}}$ is the symbol vector after the matched filter. The
entire receiver DSP chain can then be written as 
\begin{align}
	\hat{\vect{s}} = e^{-\imag \hat{\phi}} \mat{M}
	\vect{f}_\theta(\vect{r}),
\end{align}
where $\mat{M} \in \mathbb{R}^{\Nsym \times n}$ is a circulant matrix
that represents the matched filter and downsampling by a factor
$\OSd$.

\begin{remark}
Since the matched filter is linear, one could also include it in the
LDBP model as a linear layer, as was done in \cite{Haeger2018ofc}.
\RevB{Note that the phase-offset correction is genie-aided, in the sense that knowledge about the transmitted symbol vector is assumed.
Alternatively}, the phase-offset rotation could be included in the
machine-learning model where $\hat{\phi}$ would then become an
additional parameter. In this paper, these algorithms are instead
applied as separate receiver DSP block after LDBP.
\end{remark}

\RevA{\begin{remark}
	Similar to comparable prior work on complexity-reduced DBP (e.g., \cite{Ip2008, Du2010}), our system model does not include various hardware impairments and other imperfections that may arise in practice. 
	However, it has been shown that LDBP can be successfully trained in the presence of such impairments. 
	For example, \cite{Bitachon2020, Bitachon2020b} propose a training method which is blind to timing error, state of polarization rotation, frequency offset, and phase offset. 
	The approach works by first estimating these parameters using a conventional DSP chain and treating the corresponding compensation blocks as static layers during the training. 
	Similar approaches were also proposed in \cite{Oliari2020, Fan2020} to allow for the training with experimental data sets. 
\end{remark}}

\subsection{Joint Parameter Optimization}
\label{sec:joint_parameter_optimization}

In order to optimize all parameters in $\theta$, supervised learning
is applied as described in Sec.~\ref{sec:supervised_l}. The training
is performed using the Adam optimizer \cite{Kingma2015}, which is a
variant of SGD. The chosen loss function is the mean squared error
(MSE) between the correct and estimated symbols, i.e.,
$\loss(\hat{\vect{s}}, \vect{s}) = \| \hat{\vect{s}} - \vect{s} \|^2 /
\Nsym$,
where $\| \vect{s} \|^2 \define \sum_{i=1}^{\Nsym} | s_i |^2$. The
empirical loss in \eqref{eq:loss} then becomes a Monte Carlo
approximation of the expected MSE 
\begin{align}
	\mathbb{E}\left[\loss(\hat{\vect{S}}, \vect{S})\right] =
	\frac{1}{\Nsym}
	\mathbb{E}\left[\|
	\hat{\vect{S}} - \vect{S} \|^2 \right], 
\end{align}
where the expectation is over the transmitted symbols and the noise
generated by the optical amplifiers. Note that minimizing the MSE is
equivalent to maximizing the effective SNR
\begin{align}
	\label{eq:effective_snr}
	\text{SNR} = \Nsym \mathbb{E} \left[ (\| \hat{\vect{S}} - \vect{S}
	\|^2)^{-1} \right]. 
\end{align}
The effective SNR is used as the figure of merit for the numerical
results presented in the next section. 

\RevB{For a fixed link setup, our numerical results show that the optimal parameter vector depends on the input power $P$. Thus, to achieve the best performance one needs to perform a separate
optimization for each input power. 
In order to reduce the training time, we instead perform the optimization over a range of different input powers. 
More precisely, the particular input power for each transmitted frame is chosen uniformly at random from a discrete set of input powers $\mathcal{P}$.}
The set $\mathcal{P}$ is chosen based on the particular scenario.  
The power dependence of the learned solutions will be further investigated in Sec.~\ref{sec:dependence}.

\subsection{Parameter Initialization and Filter Pruning}
\label{sec:pruning}

Since SGD is a local search method, choosing a suitable
parameter-initialization scheme is important to facilitate successful
training. Our approach leverages the fact that the proposed model is
based on a well-established numerical method by initializing the
parameters such that the initial performance is close to the standard
SSM. We also experimented with other initializations, see
Sec.~\ref{sec:initialization}. 

Our parameter initialization starts by specifying the desired number
of SSM steps and this changes the number of layers in the LDBP
model. For the asymmetric SSM, the number of steps is directly
equivalent to the number of layers, i.e., $\ell = M$. For the
symmetric SSM, we have $\ell = M+1$ and the last model layer is a linear
half-step without nonlinearity as described in
Sec.~\ref{sec:generalization}. To compute the step sizes for a given
number of steps per span (StPS), the logarithmic heuristic in
\cite[Eq.~(2)]{Zhang2013c} is used with the recommended adjusting
factor $0.4$. This leads to a set of step sizes $\delta_1, \dots,
\delta_{M}$. After combining adjacent half-steps, one obtains the
lengths $\delta'_1, \dots, \delta'_\ell$, which are then used for the
filter initialization in each step. 

\begin{example}
	\RevB{For illustration purposes, let} $\Lsp = 100$ km and $\Nsp = 2$, and assume that LDBP is based on the symmetric SSM with $2$ StPS. 
	\RevB{With these assumptions, the} logarithmic step-size
	heuristic in \cite{Zhang2013c} leads to $\delta_1 = \delta_3
	\approx 70$ km and $\delta_2 = \delta_4 \approx 30\,$km.  After
	combining adjacent half-steps, the LDBP model has $\ell = 5$ layers
	and the equivalent step sizes to initialize the FIR filters are
	$\delta'_1 = 35$ km, $\delta'_2  = \delta'_3 = \delta'_4 = 50$ km,
	$\delta'_5 = 15$ km. In this case, the initial filter coefficients
	for $\vect{h}^{(2)}$, $\vect{h}^{(3)}$, and $\vect{h}^{(4)}$ will
	be the same. \demo
\end{example}

Based on $\delta'_1, \dots, \delta'_M$, a least-squares approach is
then used for computing the initial filter coefficients.\footnote{Our
previous work explored different initializations, in particular direct
truncation in \cite{Haeger2018ofc} and a multi-objective least-squares
fitting routine in \cite{Haeger2018isit}. Neither of these are
recommended anymore, see also our remarks in
Sec.~\ref{sec:multi-objective}.} Recall that the ideal frequency
response to invert CD over distance $\delta_i'$ is $H(\omega) =
e^{\imag \xi \omega^2}$, where $\xi \define - \beta_2 \delta_i' \fs^2
/ 2 $ and $\omega \define 2 \pi f / \fs$. Now, let
$\mathcal{F}(\vect{h}^{(i)}) = \sum_{k=-K_i}^{K_i} h_k^{(i)} e^{-
\imag k \omega}$ be the discrete-time Fourier transform of
$\vect{h}^{(i)}$. The standard approach is to match
$\mathcal{F}(\vect{h}^{(i)})$ to $e^{\imag \NoRev{\xi} \omega^2}$ by
minimizing the frequency-response error. After discretizing the
problem with $\omega_i = 2 \pi i / N$ for $i = -N/2, \dots, N/2$, one
obtains $\min_{\vect{h}^{(i)}}\| \vect{B} \vect{h}^{(i)} - \vect{d}
\|^2$, where $\vect{d} = (d_{-N/2}, \dots, d_0, \dots,
d_{N/2})^\transpose$ with $d_i \define e^{\imag \NoRev{\xi} \omega_i^2}$
and $\vect{B}$ is an $(N+1) \times \flen_i$ DFT matrix. In this paper,
we use a variation proposed in \cite{Sheikh2016}, where the
frequency-response error is minimized only within the signal bandwidth
and the out-of-band filter gain is further constrained to be below
some fixed maximum value, see \cite{Sheikh2016} for more details. This
procedure is applied for each of the $M$ filters.

In order to reduce the complexity of the final LDBP model, a simple
pruning approach is applied where the filters are successively
shortened during the gradient-descent optimization by forcing their
outermost taps to zero. More precisely, we start with initial filter
lengths $\flen_1' > \flen_1, \ldots, \NoRev{\flen_\ModSteps' >
\flen_\ModSteps}$ which are chosen large enough to ensure good
starting performance. The target lengths $\flen_1, \dots,
\flen_\ModSteps$ implicitly define the filter taps that need to be
removed. In our implementation, the affected taps are then multiplied
with a binary mask, where all mask values are initially $1$. At
certain predefined training iterations, some of the values are then
changed to $0$, which effectively removes the corresponding filter
taps from the model. Since all filters are assumed to be symmetric,
the pruning is done such that each pruning step always removes the two
outermost taps of a given filter (both having the same coefficient).

\begin{example}
	For the model in Example 1, assume that the initial filters have
	lengths $\flen_1' = \cdots = \flen_5' = 11$ and that the target
	lengths are set to $\flen_1 = \cdots = \flen_5 = 7$. Thus, the
	total number of pruning steps is $10$. Assuming, e.g., $100$ total
	SGD iterations, one filter is pruned every $10$ iterations, on
	average. \demo
\end{example}

The above approach does not take into account the importance of the
filter taps before pruning them. Nonetheless, we found that it leads
to both compact and well-performing models. Moreover, the final
effective SNR is relatively insensitive to the pruning details (e.g.,
which filter tap is pruned in which iteration), as long as the pruning
steps are sufficiently spread out and the target lengths $\flen_i$ are
not too small. We also found empirically that it is generally
beneficial to prune more taps in the beginning of the optimization
procedure, rather than to spread out the pruning steps uniformly. 

\RevB{\begin{remark}
	In the machine-learning literature, pruning is commonly used to compress a large model (e.g., an image classifier) in order to facilitate deployment (e.g., on mobile devices). 
	Similarly, our work shows that there exist compact solutions that could subsequently be deployed in a real-time DSP without performing any real-time optimization or pruning. 
	We have previously studied LDBP from an ASIC implementation perspective in \cite{Fougstedt2018ecoc}, where our particular design assumes that all tap coefficients are fully reconfigurable (but not trainable). 
	Thus, in terms of operation in a real scenario, one option would be to train ``offline'' a set of different filter configurations for a variety of standard fiber parameters and propagation distances. 
	Afterwards, an ``online'' testing phase could be used to identify the best-performing filters.
	To avoid large look-up tables and reduce the associated storage complexity, a conventional function approximator (e.g., a NN) could further be trained (also offline) to learn a smooth dependence between the system parameters (e.g., the fiber length) and the optimized filter coefficients. 
	If more complexity is allowed, one could also try to explore the possibility of fine-tuning a subset of the filter coefficient via implementation of a real-time gradient-descent procedure. 
	This would account for potential mismatches between the best pretrained filter coefficients and the actual optimal ones for the system under consideration.
	Assuming that the pretrained (i.e., already pruned) solutions provide good starting points for the optimization, no further real-time pruning would be required. 
\end{remark}}

\section{Numerical Results}
\label{sec:results}

\ifShowText

In this section, extensive numerical results\footnote{Source code is
available at {\tt \url{www.github.com/chaeger/LDBP}}.} are presented
in order to verify the proposed approach. \RevA{The numerical value of all system and training parameters can be found in Table~\ref{tab:parameters}.} The conventional asymmetric SSM with $500$ logarithmic StPS is used to simulate forward propagation. 
\RevA{To estimate the effective SNR after training, we average over at least $10^7$ symbols.}

\newcommand{\rr}[2]{%
    \begin{tabular}{@{}c@{}}#1 \\ #2\end{tabular}
}%

\newcommand{\FigA}{(Figs.~\ref{fig:results_Ip2008}, \ref{fig:results})}
\newcommand{\FigB}{(Figs.~\ref{fig:results_32Gbaud}, \ref{fig:results_32Gbaud_wdm})}

\setlength\tabcolsep{3.5pt} 
\newcommand{\tablehighlight}[1]{{#1}}
\begin{table}
	\caption{\RevA{List of symbols and system parameters.}}
	\label{tab:parameters}
	\centering
	\renewcommand{\arraystretch}{1.2}
	\begin{tabular}{ccc}
		\toprule
		symb. & parameter & comment / numerical value \\
		\midrule
		$\alpha$ & fiber loss parameter & $0.2$ dB/km \\
		$\mat{b}^{(i)}$ & NN bias vector &   \\
		$\beta_2$ & CD coefficient & $-21.683\,$ps$^2$/km \\
		$|\mathcal{B}_t|$ & mini-batch size& $50$ \\
		$\Delta B$ & low-pass bandwidth & $\fs$ ($37.5$ GHz in Fig.~\ref{fig:results_32Gbaud_wdm})  \\
		$\mathbb{C}$ & set of complex numbers &\\
		$\delta_i / \delta_i'$ & SSM step sizes &  \\
		$\mu$ & learning rate (Adam) & $0.001$  \\
		$\fs$ & sampling frequency &  $\OSd \fsym$ \\ 
		$\mathcal{F}$ & discrete-time Fourier trans.&   \\ 
		$\gamma$ & Kerr coefficient & $1.3$ rad/W/km \\ 
		$h$ & Planck's constant & $6.626 \cdot 10^{-34}$ Js \\ 
		$\vect{h}^{(i)}$ & LDBP filter coeffs. & \\ 
		$\vect{\eta}^{(i)}$ & ESSM filter coeffs. & \\ 
		$\kappa / \kappa_i$ & nonlinear filter param. & $20$ (Fig.~\ref{fig:results_Ip2008}) \\
		$\Lsp$ & span length & \rr{$80\,$km \FigA,}{$100\,$km \FigB} \\
		$l$ & loss function & mean squared error (MSE) \\
		$\ell$ & total LDBP steps/layers & $4$, $25$ \FigA, $21$, $41$ \FigB $\!\!\!\!$ \\
		$L$ & transmission distance & $\Nsp \Lsp$ \\
		$\Leff$ & eff.~nonlinear length &  \\
		$\mathcal{L}$ & empirical loss & \\
		$M$ & SSM steps & $500$ StPS (fwd.~propagation) \\
		$n$ & samples per frame/block & $\OSd \Nsym$ \\
		$\Nsp$ & number of spans & $25$ (Figs.~\ref{fig:results_Ip2008}, \ref{fig:results}), $10$ \FigB \\
		$\Nsym$ & symbols per frame & $1024$ (Figs.~\ref{fig:results_Ip2008}, \ref{fig:results}, \ref{fig:results_32Gbaud}), $4096$ (Fig.~\ref{fig:results_32Gbaud_wdm})  \\
		$\Nf$ & noise figure & $5$ dB \\
		$n_{\text{sp}}$ & spont.~emmission factor & $\Nf (2 (1 - e^{-\alpha \Lsp}))^{-1}$ \\
		$\nu_s$ & carrier frequency & $1.946 \cdot 10^{14} $ Hz \\
		$P$ & transmit power & $-12\,$dBm to $8\,$dBm  \\
		$\mathcal{P}$ & power set for training & scenario-dependent (see text)  \\
		$p(t)$ & pulse shape & root-raised cosine (10\% roll-off) \\
		$\OSd$ & digital oversampling & $2$ \\ 
		$\OSa$ & analog oversampling & $6$ (Figs.~\ref{fig:results_Ip2008}, \ref{fig:results}, \ref{fig:results_32Gbaud}), $10$ (Fig.~\ref{fig:results_32Gbaud_wdm})\\
		$\rho / \bm{\rho}$ & nonlinearity & time-discretized NLSE \eqref{eq:discretized_nlse} \\
		$\fsym$ & symbol rate & \rr{$10.7$ Gbd \FigA,}{ $32$ Gbd \FigB} \\
		$\vect{r}$ & observation vector & \\
		$\mathbb{R}$ & set of real numbers & \\
		$\vect{s}$ & i.i.d.~symbol vector & Gaussian (except Fig.~\ref{fig:results}(b))  \\
		$\hat{\vect{s}}$ & estimated symbol vector & \\
		$\tilde{\vect{s}}$ & symbol vector after MF & \\
		$\sigma_i / \bm{\sigma}_i$ & LDBP nonlinearity & $\bm{\sigma}_i = \bm{\sigma}_{\delta_i}$ (SSM nonlinearity)\\
		$T_i'$ & filter len.~before pruning & scenario-dependent (see text) \\
		$T_i$ & filter len.~after pruning & scenario-dependent (see text) \\
		$\theta$ & parameter vector & \\
		$u / \vect{u}$ & optical signal & \\
		$\mat{W}^{(i)}$ & NN weight matrices &   \\
		$\omega / \omega_k$ & angular frequency &   \\
		$\xi$ & scaled CD coeff. & $- \beta_2 \delta_i' \fs^2 / 2 $ \\
		$x(t)$ & transmitted signal &   \\
		$y(t)$ & received signal &   \\
		$\mathbb{Z}$ & set of integers & \\
		\bottomrule
	\end{tabular}
\end{table}

\subsection{Revisiting Ip and Kahn, 2008}
\label{sec:revisiting_Ip2008}

\begin{figure}[t]
	\centering
		\includegraphics{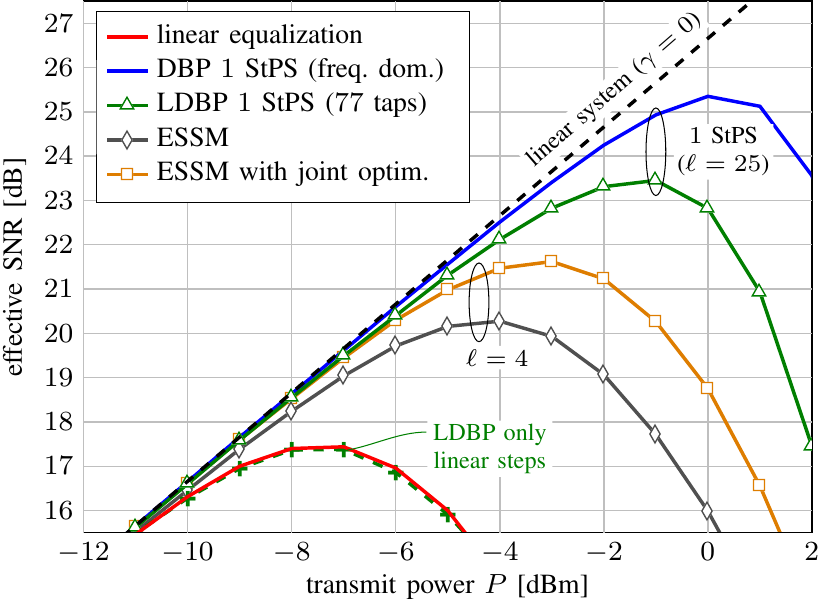}
	\caption{Results for $10.7\,$Gbaud over
	$25 \times 80\,$km fiber (similar to \cite{Ip2008}). }
	\label{fig:results_Ip2008}
\end{figure}

We start by revisiting \cite{Ip2008}, which considers single-channel
transmission of a $10.7$-Gbaud signal over $25 \times 80\,$km fiber.\footnote{\RevB{Note that compared to our simulation setup, \cite{Ip2008} assumes QPSK modulation and a Butterworth low-pass filter at the receiver.}}
For this scenario, LDBP is based on the asymmetric SSM with $1$ StPS,
i.e., $\ModSteps=25$, and pruning is performed \RevA{over $\MyNum{150000}$ gradient-descent iterations} such \NoRev{that} the final model
alternates $5$-tap and $3$-tap filters, where initially $T_i' = 9$ for
all $i$. The achieved effective SNR after training with $\mathcal{P}=
\{-2,-1,0,1\}\,$dBm is shown in Fig.~\ref{fig:results_Ip2008} by the
green triangles. As a reference, the performance of linear
equalization (red) and standard $1$-StPS DBP using frequency-domain
filtering (blue) are shown. LDBP achieves a peak SNR of $23.5\,$dB
using an overall impulse response length (defined as the length of the
filter obtained by convolving all subfilters) of $13 \cdot 4 + 12
\cdot 2 + 1 = 77$ total taps. The peak-SNR penalty compared to DBP is
around $2\,$dB, which allows us to directly compare to the results
presented in \cite{Ip2008}. Indeed, \cite{Ip2008} also considers
filter design for $1$-StPS DBP with the intention of quantifying the
complexity overhead compared to linear equalization assuming real-time
processing. Their filter design is based on frequency-domain sampling
and the same filter is used in each step. With these assumptions, it
was shown that $70$ taps/step are required to achieve performance
within $2\,$dB of frequency-domain DBP---over $20$ times more than the
number of taps required by LDBP.

\subsection{Performance--Complexity Trade-off}

To further put the above results into perspective, we perform a
similar complexity analysis as in \cite{Ip2008}, using real
multiplications (RMs) as a surrogate. For the nonlinear steps, the
exponential function is assumed to be implemented with a look-up
table. It then remains to square each sample ($2$ RMs), multiply by
$\gamma_i$ ($1$ RM), and compute the phase rotation ($4$ RMs).  This
gives $25 \cdot 7 = 175$ RMs per sample. For the linear steps, one has
to account for $13$ filters with $5$ taps and $12$ filters with $3$
taps. All filters have symmetric coefficients and can be be
implemented using a folded structure with $h_0$-normalization as shown
in \cite[Fig.~5]{Haeger2018isit}. This gives $39 \cdot 4 = 156$ RMs
per sample. In comparison, the fractionally-spaced linear equalizer in
\cite{Ip2008} requires $188$ RMs per data symbol operating at $3/2$
samples/symbol. Thus, LDBP requires $3.5$ times more RMs per symbol.
For the same oversampling factor as LDBP, a linear equalizer requires
around $69$ taps (see below). This leads to $35 \cdot 4 = 140$ RMs
with a folded implementation and a complexity overhead for LDBP of
only around $2$. If the linear equalizer is implemented in the
frequency domain, the number of RMs is reduced to $n(4\log_2 n +
4)/(n-69) \approx 50$ per sample (see, e.g.,
\cite[Sec.~4]{Secondini2016}), which increases the estimated
complexity overhead factor to $6$.  Still, these estimates are orders
of magnitude lower than the complexity overhead in \cite{Ip2008},
which was estimated to be over $100$. For a more accurate estimation
of the required hardware complexity, we refer the interested reader to
\cite{Fougstedt2018ecoc}, where LDBP is studied from an ASIC
implementation perspective including finite-precision aspects.

\begin{figure}[t]
	\centering 
	\includegraphics{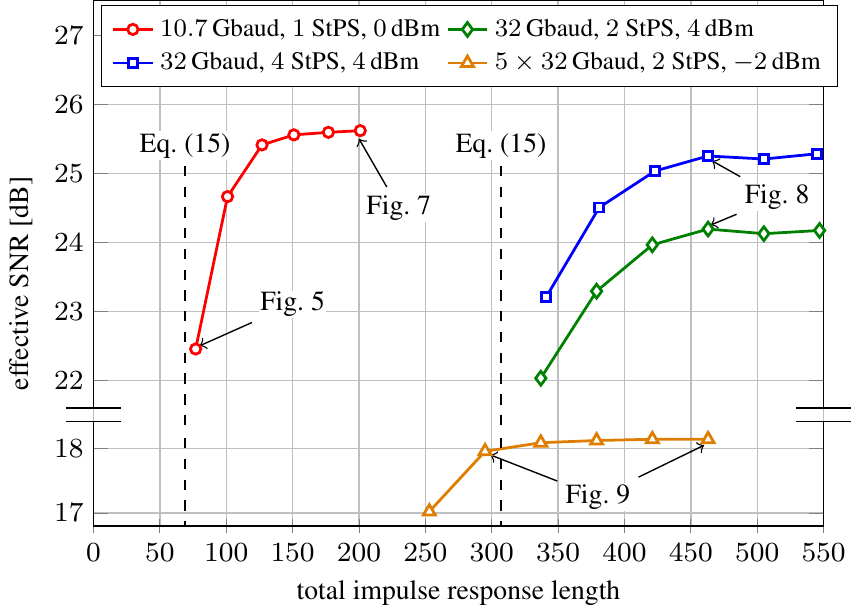}
	\vspace*{-0.15cm}
	\caption{Performance as a function of the total LDBP impulse response length.}
	\label{fig:required_length}
\end{figure}

In general, the complexity of the linear steps scales directly with
the number of filter taps. The red line in
Fig.~\ref{fig:required_length} shows the effective SNR of LDBP at $P =
0\,$dBm as a function of the overall impulse response length at
various levels of pruning. The right-most point of the curve
corresponds to $T_i = T_i' = 9$ for all $i$, i.e., no pruning. 
\RevA{To generate each curve in Fig.~\ref{fig:required_length}, we progressively prune the model for a total of $\MyNum{150000}$ gradient-descent iterations, and save the partially pruned model after every $\MyNum{25000}$ iterations.}
It can
be seen that by allowing more than $77$ total taps, performance
quickly increases up to around $25.5\,$dB, which is slightly higher
than frequency-domain DBP, see Fig.~\ref{fig:results_Ip2008}. As a
reference, the dashed line in Fig.~\ref{fig:required_length} indicates
the required number of filter taps that can be expected based on the
memory that is introduced by CD. To estimate the memory, one may use
the fact that CD leads to a group delay difference of $2 \pi |\beta_2|
\Delta f L$ over bandwidth $\Delta f$ and transmission distance $L$.
Normalizing by the sampling interval $\fs^{-1}$, this confines the
memory to approximately 
\begin{align}
	\label{eq:required_length}
	T_\text{cd} = 2 \pi \beta_2 \Delta f L \fs
\end{align}
samples. The bandwidth $\Delta f$ depends on the baud rate, the pulse
shaping filter and the amount of spectral broadening. For a
conservative estimate, we use $\Delta f = (1 + 0.1) \cdot 10.7\,$GHz
(i.e., spectral broadening is ignored) which leads to $T_\text{cd}
\approx 69$.

\begin{figure*}[t]
	\centering
		\fcolorbox{white}{white}{%
		\subcaptionbox{}{\includegraphics{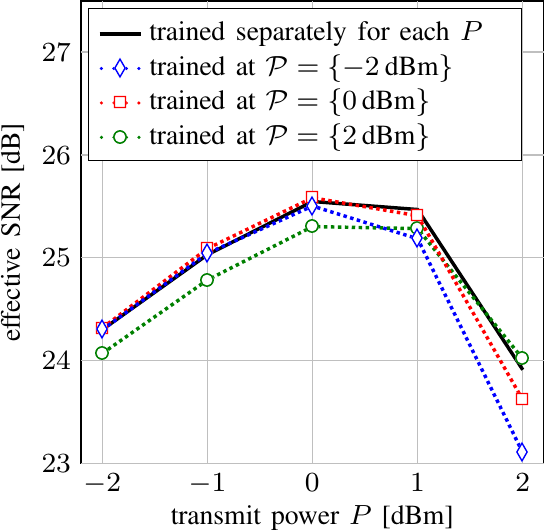}}
		\subcaptionbox{}{\includegraphics{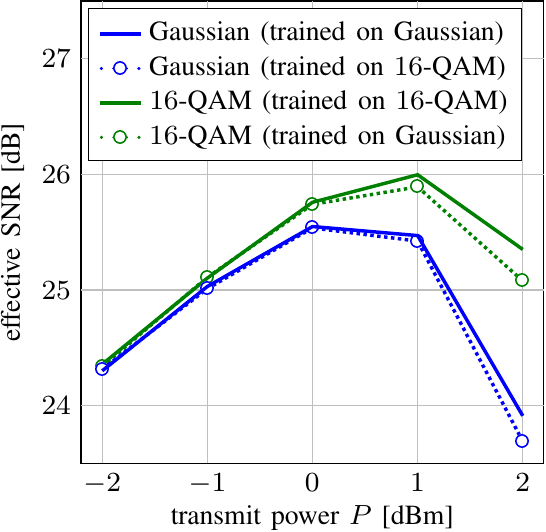}}
		\subcaptionbox{}{\includegraphics{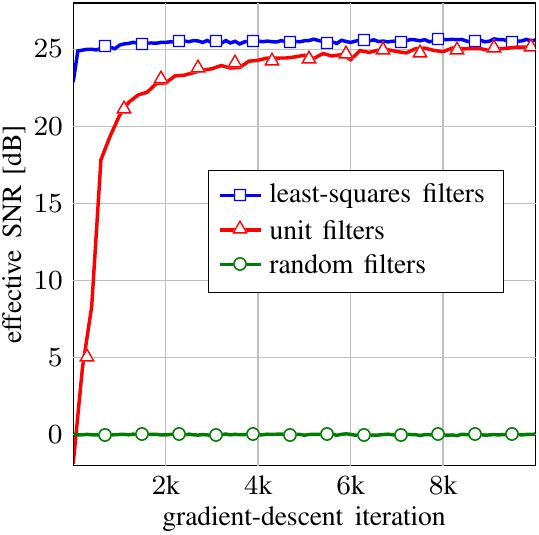}}
		}%
		\caption{Additional results for $10.7\,$Gbaud over
		$25 \times 80\,$km fiber: (a) power dependence, (b)
		modulation-format dependence, (c) parameter-initialization
		schemes.}
	\label{fig:results}
\end{figure*}

\subsection{Improving the Enhanced SSM (Filtered DBP)}
\label{sec:filtered_dbp}

We now consider the ESSM which includes filters in the nonlinear
steps. We start with the conventional ESSM, where a single filter is
optimized (starting from a unit filter) and then applied in each step.
Excluding the overhead due to overlap-and-save techniques, one ESSM
step requires $4 \log_2 n + 11 + \kappa$ RMs per sample \cite[Sec.~4,
single pol.]{Secondini2016}, where we recall that $2 \kappa + 1$ is
the length of the filter in the nonlinear steps. We perform $4$ steps
with $\kappa = 20$, which gives roughly the same number of RMs as the
$77$-tap LDBP model. The performance \RevA{after training for $5000$ iterations} is shown by the grey diamonds in
Fig.~\ref{fig:results_Ip2008}. The ESSM achieves a smaller peak SNR by
around $3\,$dB compared to LDBP, highlighting the advantage of using
more steps. With the proposed modification of jointly optimizing all
filters and assuming the same filter length as before (i.e., $\kappa_1
= \ldots = \kappa_4 = 20$), more than $1\,$dB performance improvements
over the conventional ESSM can be obtained \RevA{for the same number of training iterations} (yellow squares in
Fig.~\ref{fig:results_Ip2008}).

\subsection{Generalization Error}
\label{sec:dependence}

Next, we study to what extend the learned solutions generalize beyond
the training data, focusing on the transmit power and the employed
modulation format. To do this efficiently, we consider the same setup
as before, albeit using longer filters ($9$ taps/step) and without
performing any pruning. \RevA{In all cases, we train for $1500$ iterations}. Fig.~\ref{fig:results} (a) shows the
performance after training separately for each input power $P$ (solid
black line), compared to the case where training is performed at a
specific power and the learned solution is then used at the other
powers without retraining. From these results, it is clear that the
optimal parameters are power-dependent in general and that the
generalization error increases with the distance between the training
and testing power. As mentioned in
Sec.~\ref{sec:joint_parameter_optimization}, our general approach to
achieve a compromise between performance and training time is to
optimize over a set of powers $\mathcal{P}$. The set $\mathcal{P}$
corresponds approximately to the power region where the performance
peak (in terms of effective SNR) is expected. It is important to
stress, however, that training over a large power range is not
recommended, since the loss function in this case is dominated by
regions with low effective SNR (i.e., large MSE). 

Fig.~\ref{fig:results} (b) shows the case where the modulation format
differs between training and testing. Here, we consider the case where
training is performed with Gaussian symbols and the solution is
applied to $16$-QAM (and vice versa). In this case, LDBP appears to
generalize well beyond the training data, especially in the linear
operating regime.

\subsection{Parameter Initialization}
\label{sec:initialization}

As explained in Sec.~\ref{sec:pruning}, the filter coefficients are
carefully initialized using least-squares fitting to the (per-step)
inverse CD response. In general, we found that the chosen parameter
initialization plays a critical role in determining the optimization
behavior. To illustrate this, Fig.~\ref{fig:results} (c) shows the
learning curves obtained for the same scenario as in
Sec.~\ref{sec:dependence}, trained at $\mathcal{P} =
\{0\,\text{dBm}\}$ with three different initialization schemes:
least-squares filters, unit filters, and random filters. For the
random case, the real and imaginary part of all taps are
i.i.d.~$\mathcal{N}(0,1)$. All filters are then normalized according
to $\vect{h}^{(i)} \leftarrow \vect{h}^{(i)} / \sum_{j=-T_i}^{T_i}
|h_j^{(i)}|^2$ for $i = 1, \ldots, M$, which avoids numerical
instabilities. However, the optimization remains essentially ``stuck''
at $0\,$dBm without improvements, suggesting the existence of a
barrier in the optimization landscape that is difficult to overcome.
\RevC{This is somewhat surprising, given the fact that Gaussian initialization is widely used in the machine-learning literature.}
Unit filters converge to a reasonable SNR, albeit at a much slower
convergence speed than the least-squares filters. Even though the
final SNR after $10$k iterations is not quite competitive ($25.1\,$dB
vs.~$25.6\,$dB for least squares), this initialization has the
advantage of being agnostic to the link setup, i.e., it does not
require knowledge about parameters such as the fiber length or the
dispersion coefficient.

\subsection{Higher Baud Rates}

We now move on to higher baud rates and consider single-channel
transmission of $32\,$Gbaud over $10 \times 100\,$km of fiber. For
this case, LDBP is based on the symmetric SSM with logarithmic step
sizes. We use $2$ StPS and $4$ StPS, leading to $\ell = 21$ and $\ell
= 41$, respectively. We start by investigating the number of filter
taps required to achieve good performance. Results for $\mathcal{P} =
\{4\,\text{dBm}\}$ are shown in Fig.~\ref{fig:required_length} by the
green ($2$ StPS) and blue ($4$ StPS) lines, \RevA{assuming again $\MyNum{150000}$ total iterations and saving the partially pruned models after every $\MyNum{25000}$ iterations}. As a reference,
\eqref{eq:required_length} with $\Delta f = (1+0.1) \cdot 32\,$GHz, $L
= 1000\,$km, and $\fs = 64\,$GHz gives $T_\text{cd} = 307$ taps, which
is shown by the dashed line. Similar to before, the performance drops
sharply when the filters are pruned too much and the impulse response
length approaches the CD memory. For comparison, we note that Martins
et al.~consider filter design (and quantization) for time-domain DBP
assuming $32\,$Gbaud signals in \cite{Martins2018}. Their filters
require $301$ taps to account for one span of $108\,$km fiber.
Extrapolating to $L = 1000\,$km and adjusting for the slighty
increased dispersion parameter in \cite{Martins2018} ($20.17\,$
ps/nm/km vs $17\,$ ps/nm/km), the total number of taps would be around
$301 \cdot 1000/108 \cdot 17/20.17 \approx 2350$ which is
significantly larger than the results in
Fig.~\ref{fig:required_length}. 

\begin{figure}[t]
	\centering
		\includegraphics{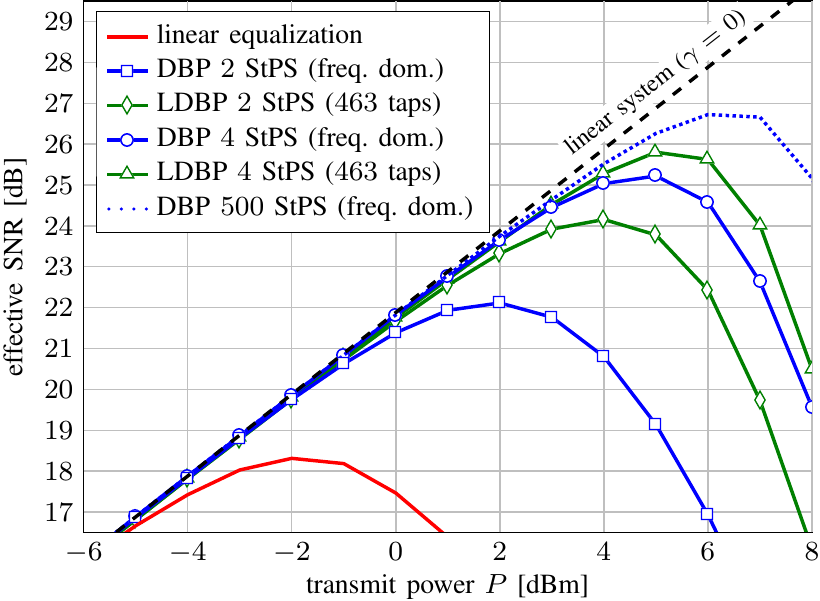}
		\caption{Results for $32\,$Gbaud over $10 \times 100\,$km fiber}
	\label{fig:results_32Gbaud}
\end{figure}

Another interesting observation from Fig.~\ref{fig:required_length} is
that for the same performance, the model with more steps requires
fewer total taps, and hence less complexity to implement all linear
steps. This is interesting because it is commonly assumed that the
complexity of the SSM/DBP simply scales lineary with the number of
steps, leading to an undesirable complexity increase. On the other
hand, our results show that by carefully optimizing and pruning the
linear steps, a very different complexity scaling can be achieved.
Indeed, doubling the number of steps reduces the expected CD memory
per step in half, which allows us to decrease the per-step complexity
by reducing the filter lengths (ideally by half). The additional
performance gain due the larger step count can then be traded off for
fewer taps, as shown in Fig.~\ref{fig:required_length}.

Fig.~\ref{fig:results_32Gbaud} shows the performance of the two LDBP
models with $463$ taps as a function of transmit power, where we
retrain the solutions obtained in Fig.~\ref{fig:results_32Gbaud} with
$\mathcal{P} = \{3,4,5\}\,$dBm ($2$ StPS) and $\mathcal{P} =
\{4,5,6\}\,$dBm ($4$ StPS) \RevA{for an additional $1500$ steps}. Significant performance improvements are
obtained compared to DBP with the the same number of StPS. In
particular, the peak SNR is increased by $2.1\,$dB ($2$ StPS) and
$0.6\,$dB ($4$ StPS), respectively. In the light of Remark
\ref{rmk:special_cases}, we conjecture that the gains can be partially
attributed to the implicit joint optimization of step sizes and
nonlinear operator placements.

\subsection{Wavelength Division Multiplexing}

\begin{figure}[t]
	\centering
		\includegraphics{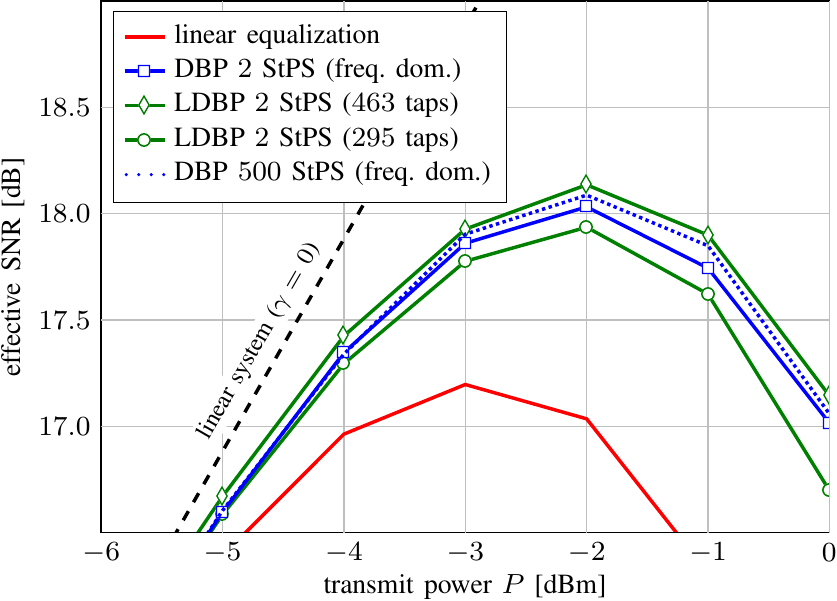}
		\caption{Results for $5 \times 32\,$Gbaud (WDM) over $10 \times 100\,$km fiber}
	\label{fig:results_32Gbaud_wdm}
\end{figure}

Lastly, we study WDM transmissions, where it is known that the
potential performance gains provided by static single-channel
equalizers are limited due to nonlinear interference from neighboring
channels, see, e.g., \cite{Dar2017}.\footnote{In general, adaptive
equalization can provide more gains due to the time-varying nature of
the interference.} On the other hand, the resulting accuracy
requirements for solving the NLSE are significantly relaxed compared
to single-channel transmission because of the lower achievable
effective SNRs. This provides an opportunity to further reduce
complexity of LDBP by pruning additional filter taps. To demonstrate
this, we consider $5 \times 32\,$Gbaud WDM channels, where the channel
spacing is $37.5\,$GHz. Compared to before, the receiver low-pass
bandwidth is reduced to $\Delta B = 37.5\,$GHz to filter out the
center channel and the oversampling factor to simulate forward
propagation as well as the number of symbols per block are increased
to $\OSa = 10$ samples/symbols and $\Nsym = 4096$, respectively.
Fig.~\ref{fig:results_32Gbaud_wdm} shows the performance of the
$2$-StPS LDBP model with $463$ taps from
Fig.~\ref{fig:results_32Gbaud}, retrained for this scenario with
$\mathcal{P} = \{-3,-2,-1\}\,$dBm (green diamonds) \RevA{for $1500$ steps}. This model already
outperforms ``ideal'' DBP\footnote{\RevA{A possible explanation for this performance improvement is provided in \cite{Fan2020}, where the authors argue that the learned filter coefficients try to strike a balance between inverting the channel and minimizing additional nonlinear phase-noise distortions due to noise-corrupted signal power levels.}} with $500$ StPS and increasing the number of
steps or taps does not lead to significant improvements. On the other
hand, further pruning the model by $36$\% to $295$ taps only gave a
peak-SNR penalty of around $0.2\,$dB (green circles), see also the
yellow triangles in Fig.~\ref{fig:required_length}.

\begin{figure*}[t]
	\centering
		\subcaptionbox{Conventional approach}{\includegraphics{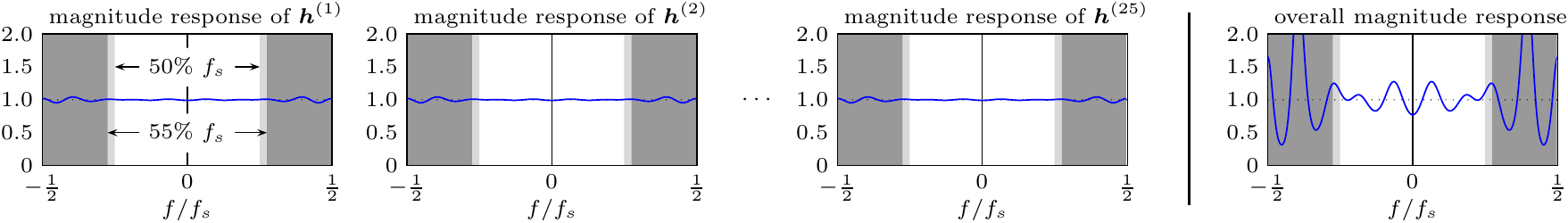}}

		\vspace{0.3cm}

		\subcaptionbox{Learned approach}{\includegraphics{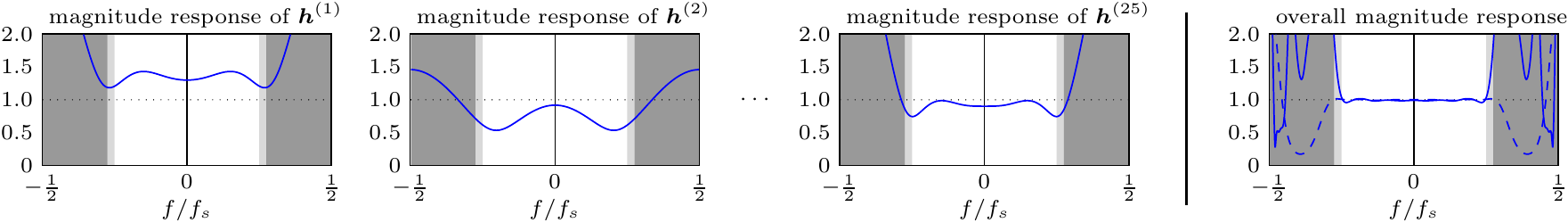}}

		\caption{Per-step filter responses (left) and overall filter
		response (right) for the simulation scenario in \cite{Ip2008}
		at $P = -1\,$dBm, see Sec.~\ref{sec:revisiting_Ip2008}.}
	\label{fig:response}
\end{figure*}

\fi

\section{Examining The Learned Solutions}
\label{sec:examining}

\ifShowText

In this section, some of the obtained numerical results will be
examined in more detail with a focus on the question ``What can be
learned from the resulting systems?''. Our goal is to explain why the
filters in LDBP can be pruned to such short lengths compared to prior
work, while still maintaining good performance. Indeed, we were
pleasantly surprised that the machine-learning approach uncovered a
simple but effective design strategy that had not been considered
earlier. We also provide additional theoretical justifications and
interpretations for the learned parameter configurations.

\subsection{Conventional Filter Design for Digital Backpropagation}

Filter design for the SSM has been considered in multiple prior works,
both for enabling real-time DBP \cite{Ip2008, Zhu2009, Goldfarb2009,
Fougstedt2017, Fougstedt2017b, Fougstedt2018ptl, Martins2018} and as a
means to lower the simulation time of the forward propagation process
\cite{Li2005a, Farhoudi2010, Zhu2012}. The standard filter-design
approach is to optimize a single filter and then use it repeatedly in
each step. To compute the individual filter coefficients, a wide
variety of approaches have been proposed. For instance, since the
inverse Fourier transform of the CD response $H(\omega) = e^{\imag \xi
\omega^2}$ can be computed analytically, filter coefficients may be
obtained through direct sampling and truncation \cite{Savory2008}.
Other approaches include frequency-domain sampling \cite{Ip2008},
wavelets \cite{Goldfarb2009}, truncating the inverse DFT response
\cite{Haeger2018ofc}, and least squares \cite{Eghbali2014,
Sheikh2016}. However, all of these methods invariably introduce a
truncation error due to the finite number of taps. Since repeating the
same filter $M$ times raises the frequency response to the $M$-th
power, the standard filter-design approach therefore leads to a
coherent accumulation of truncation errors. In essence, using the same
filter many times in series magnifies any weakness. To illustrate
this, Fig.~\ref{fig:response} (a) shows an example of the per-step
responses (left) and overall response (right) for the $10.7\,$Gbaud
scenario, where a standard least-squares fitting with $15$ taps/step
is used. While the individual responses are relatively frequency-flat
over the signal bandwidth, the overall response exhibits severe
truncation errors.

The problem associated with applying the same filter multiple times in
succession is of course well known and recognized in the literature.
In fact, when commenting on their results in the book chapter
\cite{IpKahn2009}, the authors explicitly state that the $70$ taps
required for the filters in \cite{Ip2008} are ``much larger than
expected''. They use the term ``amplitude ringing'' in order to
explain this result: ``Since backpropagation \emph{requires} multiple
iterations of the linear filter, amplitude distortion due to ringing
accumulates'' \cite{IpKahn2009} (emphasis by us). The ``requirement''
mentioned in \cite{Ip2008} to repeat the same filter seems to have
implicitly guided almost all prior work on filter design for DBP. A
notable exception is \cite{Zhu2009}, where a complementary filter pair
is optimized, rather than just one filter. While this does lead to
improvements, one may argue that using the same filter pair multiple
times in succession is again suboptimal and leads to the same
``amplitude ringing'' problem.

\subsection{Learned Filter Responses}
\label{sec:multi-objective}

The learned per-step filter responses for the LDBP model alternating
$5$-tap and $3$-tap filters (see Fig.~\ref{fig:results_Ip2008}) are
shown in the left of Fig.~\ref{fig:response} (b). At first glance,
they may appear somewhat counterintuitive because these responses are
generally much worse approximations to the ``ideal'' frequency-flat CD
response, especially when compared to the filters obtained by
least-squares fitting shown in Fig.~\ref{fig:response} (a). On the
other hand, when inspecting the overall magnitude response shown in
the right of Fig.~\ref{fig:response} (b), a rather different picture
emerges. Indeed, the combined response follows an almost perfectly
frequency-flat behavior. Another interesting observation is related to
the fact that the learned overall response appears to exhibit a gain
in the spectral region corresponding to the excess bandwidth of the
pulse-shaping filter, i.e., the region between $50$\% and $55$\% of
$\fs$, shaded in light gray. One may \RevB{therefore} wonder if this gain is
of any significance, e.g., in terms of performance. Here, we argue
instead that the learned approach has identified this spectral region
as less important for the filter design, in the sense that a
potentially larger truncation error can be tolerated because this
region does not contain much of the signal power. Indeed, if one
allows for more filter taps ($9$ taps, dashed line), the learned
overall response becomes flat also in this region.

The machine-learning approach demonstrates that a precise
approximation of the exact CD response in each step is not necessary
for achieving good performance and that sacrificing individual filter
accuracy can lead to a much better overall response. A theoretical
justification for this observation can be obtained by regarding the
filter-design problem for DBP as a multi-objective optimization
problem. More precisely, instead of the standard least-squares
formulation where only individual responses are optimized to fit the
CD response, additional objectives should formulated based on the
combined responses of neighboring filters, and in particular the
overall response. In order to formalize this, consider the set of
objectives 
\begin{equation}
	\begin{aligned}
	\label{eq:objectives}
\mathcal{F}(\vect{h}^{(i)}) &\circeq e^{\imag \xi
\omega^2}, &&i = 1,2, \dots, M\\
\mathcal{F}(\vect{h}^{(i)} * \vect{h}^{(i+1)}) &\circeq e^{\imag 2
\xi
\omega^2}, &&i = 1,2, \dots, M-1\\
&\,\,\,\vdots && \\
\mathcal{F}(\vect{h}^{(1)} * \cdots * \vect{h}^{(M)}) &\circeq
e^{\imag M \xi
\omega^2},
	\end{aligned}
\end{equation}
where the symbol $\circeq$ may be interpreted as ``should be close
to''. The conventional filter-design approach only utilizes the first
objective based on the individual responses, in which case one finds
that all filters should be the same. However, it can be shown that
better filters are obtained by solving an optimization problem that
targets all of the above objectives. In particular, keeping the
coefficients for all but one filter constant, \eqref{eq:objectives}
can be written as a standard \emph{weighted} LS problem as follows.
Since, e.g., $\mathcal{F}(\vect{h}^{(i)} * \vect{h}^{(i+1)})
=\mathcal{F}(\vect{h}^{(i)}) \mathcal{F}(\vect{h}^{(i+1)})$, we have
$(\mat{B} \vect{h}^{(i)}) \circ (\mat{B} \vect{h}^{(i+1)})$ in the
discretized problem, where $\circ$ denotes element-wise
multiplication. Hence, one obtains 
\begin{align}
	\label{eq:weighted_ls}
	\min_{\vect{h}^{(i)}} \sum_{j=1}^{O_i} \lambda_j \| (\vect{B}
	\vect{h}^{(i)}) \circ \vect{e}_j - \vect{d}_j \|^2,
\end{align}
where $O_i$ is the number of objectives, $\lambda_j > 0$ are weights,
$\vect{e}_j$ are constant vectors representing the influence of other
filters and $\vect{d}_j$ are the discretized objective vectors. This
suggests a simple strategy for the joint filter optimization by
solving the weighted LS problem for each of the $M$ filters in an
iterative fashion. The weights $\lambda_1, \dots, \lambda_{O_i}$ can
be chosen based on a suitable system criterion, e.g., effective SNR.
In \cite{Haeger2018isit}, this approach was used to find better
initializations compared to the filter truncation method used in
\cite{Haeger2018ofc}. While this approach does give improvements, it
is not recommended anymore and we instead observe consistently better
results by using the pruning approach described in
Sec.~\ref{sec:pruning}. Nonetheless, we believe that the above
formulation as a multi-objective optimization provides valuable
insights into the filter-design problem.

Lastly, we point out that the performance when using only the linear
steps in LDBP after training reverts approximately to that of a linear
equalizer, as shown by the dotted green line (crosses) in
Fig.~\ref{fig:results_Ip2008}. This leads to another intuitive
interpretation of the task that is accomplished by deep learning. In
particular, the optimized filter coefficients represent an approximate
factorization of the overall linear inverse fiber response. At first,
this may seem trivial because the linear matrix operator $e^{L
\mat{A}}$ can be factored as $e^{\delta \mat{A}}\cdot \dots \cdot
e^{\delta \mat{A}}$ with $L = \delta M$ for arbitrary $M$ to represent
shorter propagation distances. However, the factorization task becomes
nontrivial if we also require the individual operators $e^{\delta
\mat{A}}$ to be ``cheap'', i.e., implementable using short filters.

\begin{remark}
	\RevA{A natural framework to study the factorization of FIR filters is via their $z$-transform. 
	Recall that the $z$-transform of a symmetric FIR filter with $T$ taps is
\begin{align}
	H(z) = \sum_{n=0}^{T-1} h_n z^{-n}. 
\end{align}
The polynomial $H(z)$ can be factorized according to 
\begin{align}
	H(z) = h_{T-1} \prod_{i=1}^{T-1} (z^{-1}-q_i), 
\end{align}
where $q_i$ is the $i$-th root of $H(z)$. This factorization can be interpreted as splitting the original filter into a cascade of $(T-1)/2$ symmetric $3$-tap filters. 
Note that while this splitting is exact, it is not unique, in the sense that the individual $3$-tap filters can be arbitrarily rescaled while preserving the overall filter gain (e.g., multiplying all coefficients of the first filter by $2$ and dividing all coefficient of any other filter by $2$ gives the same overall response).
	We experimented with this factorization based on the impulse response of a linear time-domain equalizer for the entire propagation distance.
	However, this approach gives no control over the individual filter
	responses, other than the choice of how to distribute the overall
	gain factor. Moreover, it is not obvious how to achieve a good
	ordering of sub-filters in the SSM. 
}
	
\end{remark}

\fi

\section{Conclusions and Future Work}
\label{sec:conclusion}

We have proposed a novel machine-learning approach for fiber-optic
communication systems based on parameterizing the split-step method
for the nonlinear Schrödinger equation. The resulting physics-based
model was shown to have a similar mathematical structure compared to
standard deep neural networks. It also comes with several compelling
advantages:

\begin{itemize}
	\item \emph{Clear hyperparameter choices:} A major issue with
		standard NN models is the absence of clear guidelines for the
		network design, e.g., choosing the number of layers or the
		activation function. Our approach is based on a well-known
		numerical method where the ``activation function'' corresponds
		to the Kerr effect and one can leverage prior domain knowledge
		for selecting hyperparameters, specifically for choosing an
		appropriate number of layers.

	\item \emph{Good parameter initializations:} Deep learning relies
		on local search methods, i.e., gradient descent. In general, it
		is nontrivial to find parameter-initialization schemes that
		facilitate successful training and allow for a reliable
		convergence to a good solution. For the proposed model, it was
		shown that good starting points for the optimization can be
		obtained by initializing parameters close to the standard
		split-step method.

	\item \emph{Model Interpretability:} While neural networks are
		universal function approximators, it can be difficult to
		understand their inner workings and interpret the learned
		parameter configurations. On the other hand, the proposed model
		is based on familiar building blocks---FIR filters---where the
		learned frequency responses can reveal new theoretical insights
		into the obtained solutions.

\end{itemize}

\begin{figure}[t]
	\centering
		\includegraphics{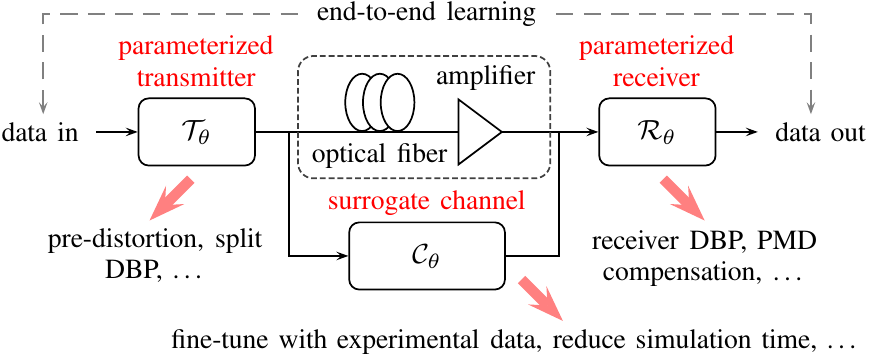}
	\caption{Potential applications of the proposed approach 
	in the context of end-to-end autoencoder learning \cite{OShea2017}. 
	$\pTX, \pRX, \pCH$ are parameterized functions.}
	\label{fig:model_based}
\end{figure}

For future work, we note that while digital backpropagation is now
most often used at the receiver, it was first studied as a transmitter
pre-distortion technique \cite{Essiambre2005, Roberts2006}. Therefore,
besides receiver-side equalization, other potentially interesting
applications include the learning of transmitter signaling schemes, or
the optimization of both transmitter \emph{and} receiver models
similar to split nonlinearity compensation \cite{Lavery2016a}, see
Fig.~\ref{fig:model_based}. The joint optimization of parameterized
transmitters and receivers would also connect our work to the recently
proposed end-to-end autoencoder learning approach in \cite{OShea2017}.
In general, this approach requires a differentiable channel model in
order to compute gradients for the transmitter optimization. Since the
proposed machine-learning model is based on the NLSE (and thus on the
optical fiber channel), it can also serve as the basis for a
fine-tuned surrogate channel, e.g., based on experimental data using
techniques similar to \cite{OShea2019, Ye2018}.

\balance

\end{document}